\documentstyle[12pt,a4,amstex,epsfig]{article}
\begin{document}

\begin{flushright}
MPI-PhT/96-14\\
March 1996\\
\end{flushright}
\begin{center}
\large {\bf QCD Corrections to W Pair Production at LEP200}\\
\mbox{ }\\
\normalsize
\vskip3cm
{\bf K.J. Abraham}
\vskip0.3cm
Dept. of Physics, University of Natal \\
Pietermaritzburg, South Africa \\
\vskip2cm
{\bf Bodo Lampe}               
\vskip0.3cm
Max Planck Institut f\"ur Physik \\
F\"ohringer Ring 6, D-80805 M\"unchen \\
\vspace{3cm}

{\bf Abstract}\\
\end{center}
One loop QCD corrections to hadronic W decay are 
calculated for arbitrary W polarizations . The results are 
applied to W pair production and decay at LEP200.  
We focus on 
the corrections to angular distributions with
particular emphasis on azimuthal
distributions and correlations.
The relevance of our results to the experimental determination of
possible nonstandard triple gauge bosons interactions is discussed.

\newpage

{\bf 1. Introduction} 

The LEP200 experiment at CERN will start data taking above the WW 
threshold in 1997. 
While much of the emphasis of the LEP200 program has been on the 
possible discovery of the Higgs and the precise determination of $m_W$, 
there
remains the important aspect of probing the triple gauge boson couplings 
which are an essential feature of the non-abelian gauge symmetry of the 
standard model. Of course, there is already indirect
evidence for the existence of triple gauge boson couplings from the high 
precision of the data at LEP1 \cite{rujula}, but extracting 
bounds on these couplings 
requires unavoidable model dependent assumptions.
Even though the 
statistical accuracy of the LEP200 data will not be as impressive as
at LEP1, it will be sufficient 
to remove any doubts concerning the 
structure of the pure gauge sector in the standard model. 

It is expected that at threshold, radiative corrections may play a 
significant role in W pair production.
Electroweak corrections have been known 
for some time \cite{veltman} and there has been some progress towards
including finite $W$ width effects, which arise, for example, 
from the interference between the 
processes $e^+e^- \rightarrow f_1 \bar f_2 f_3 \bar f_4$ with and 
without intermediate W's \cite{kleiss}. In this paper we present the 
one loop perturbative QCD corrections 
in the case where at least one of the W's decays hadronically, 
$W \rightarrow q \bar q'(g)$, which to the best of
our knowledge have never been calculated along the lines we follow
\cite{maina}.

In principle it is possible that higher order corrections modify 
the standard model predictions in such a way that they mimic
the existence of nonstandard triple gauge boson  couplings, if not taken 
into account. 
Therefore, our results should be compared to possible effects from 
nonstandard couplings as parameterised in  \cite{peccei} 
\cite{gounaris}. 
It is conceivable that nonstandard effects enter at the few percent 
level and are thus of the same order of magnitude as the 
QCD effects discussed in this paper. 

It would be possible to completely neglect higher order QCD corrections by 
studying only final states in which both the W's decay leptonically, however
much information concerning 
the triple gauge boson vertices will be inevitably
washed out due to incomplete kinematical reconstruction, quite apart from 
large statistical errors due to low event rates in this channel. Therefore 
hadronic decays are important. 
For definiteness, we shall mostly  
consider in this article the process 
$e^+e^- \rightarrow W^+W^- \rightarrow l^+\nu q\bar q'(g)$ where 
only the $W^-$ decays hadronically, although our result can easily 
be extended to 
$e^+e^- \rightarrow W^+W^- \rightarrow q_1 \bar q_2 q_3 \bar q_4$. 
Our treatment is along the lines of \cite{gounaris} modifying the 
discussion of differential distributions sensitive to non standard 
triple gauge boson vertices to include the effects of higher order QCD 
corrections. 
Let us stress once again that the W's are assumed on--shell 
throughout the paper (as in \cite{gounaris}) 
so that our result does not take care of finite 
width effects whatsoever. 

In section 2 we shall discuss "inclusive jet angular 
distributions"  (i.e. the matrix element for $W \rightarrow q \bar q'g$ 
is integrated in such a way that it adds to the lowest order 
cross section 
$e^+e^- \rightarrow W^+W^- \rightarrow l^+\nu q\bar q'$) ,   
but for completeness, we include the fully 
differential results in the appendix. 
In eq. \ref{eq20} 
we present a nice compact formula, which summarises the most  
important piece of our result.  
In section 3 we shall discuss in detail some phenomenological 
consequences of our result on various 
differential distributions. 

\vskip2cm

{\bf 2. A Detailed Description of the Calculation} 

Non expert readers might be tempted to think that QCD corrections 
to $W \rightarrow q \bar q'$ and consequently also for 
$e^+e^- \rightarrow W^+W^- \rightarrow l^+\nu q\bar q'$,
are given by a correction factor 
$(1+{\alpha_s / \pi})$. However, this is true only as far 
as the total rate is concerned, but is not the case if one considers some 
of the  
angular disributions relevant for probing triple gauge boson vertices.
The corrections to the angular distributions are related to QCD corrections 
to hadronic Z decay \cite{zerwas}, but are more complicated due to the 
larger number of particles in the final state. 
Correspondingly, more 
angles are necessary to describe the cross section. 

Following the treatment 
of \cite{gounaris} the cross-section can be written as  
\begin{equation}
{d\sigma(e^+e^- \rightarrow W^+W^- \rightarrow l^+ \nu_l 
q \bar q') \over d\cos \vartheta d\cos \theta_l d\phi_l 
d\cos \theta d\phi}
=\sum_{A,B,A',B',\lambda} F^{\lambda}_{ABA'B'}(s,\cos \vartheta) 
D^0_{AB}(\theta ,\phi)D^0_{A'B'}(\pi-\theta_l,\phi_l-\pi) 
\label{eq1}
\end{equation}
where the sum runs over the polarizations A,B=L,+,-- of the $W^-$ 
and A',B'=L,+,-- of the $W^{+}$ and the electron helicity $\lambda$. 
$D^0_{AB}$ are the "decay functions" of the decay $W^-\rightarrow q \bar q'$, 
normalized such that 
\begin{equation}
\int_{-1}^{+1}d\cos \theta \int_0^{2\pi}{d\phi \over 2\pi}
D^0_{AB}(\theta ,\phi)={4\over 3} \delta_{AB} 
\label{eq2}
\end{equation}
They depend on two angles (polar angle $\theta$ and azimuthal angle 
$\phi$) which specify the direction of the 
outgoing quark $q$ in the $W^-$ rest frame with respect to the 
direction defined by the $W^-$ motion in the lab frame. 
The definition of $\theta_l$ and $\phi_l$ is similar.  
The functions $F^{\lambda}_{ABA'B'}$ can be constructed from the 
helicity amplitudes for $e^{+}e^{-} \rightarrow W^{+}W^{-}$. 
They depend on the total 
$e^+e^-$ energy $\sqrt{s}$ and the production angle $\vartheta$ 
of the W's. Since we are interested in QCD corrections to W--decay 
we shall not discuss the functions $F^{\lambda}_{ABA'B'}$ any further. 
Explicit expressions can be found in ref. \cite{gounaris}.

For the sake of convenience we reproduce from \cite{gounaris} the lowest order 
decay functions,
\begin{equation}
\begin{split}
D_{LL}^0(\theta ,\phi)&=\sin^2\theta \\
D_{\pm\pm}^0(\theta ,\phi)&={1 \over 2}(1\pm\cos\theta)^2 \\ 
D_{+-}^0(\theta ,\phi)&={1 \over 2}\sin^2\theta e^{2i\phi} \\ 
D_{\pm L}^0(\theta ,\phi)&=
(\pm\cos\theta\sin\theta-\sin\theta) 
{ e^{\pm i\phi} \over \sqrt{2}}
\label{eq3}   
\end{split} 
\end{equation} 
These results can be derived by 
contracting the "hadron tensor" $H_{\mu\nu}(W^- \rightarrow q \bar q')$ 
with the corresponding $W$ polarization vectors $\epsilon_A^{\mu}$,   
$D_{AB}=H_{\mu\nu}\epsilon_A^{\mu}\epsilon_B^{\ast\nu}$. 
All the rest of the decay functions can be obtained by using the relation  
\begin{equation}
D_{AB}(\theta ,\phi)=D^{\ast}_{BA}(\theta ,\phi) 
\label{eq4}
\end{equation} 
which is true beyond the leading order.  
By summing up all diagonal decay functions one gets  
\begin{equation} 
D^0_{total}=2 
\label{eq5}
\end{equation} 
Alternatively, this can be obtained from 
$D_{total}=\sum_{A,B}H_{\mu\nu}(-g_{\mu\nu}+{W_{\mu}W_{\nu}\over m_W^2})$, 
where $W$ denotes the 4--momentum of the $W^-$. 
It should be stressed that throughout this work, even when 
discussing the phenomenology in section 3, $\theta$ and $\phi$ 
are defined in the rest-frame of the decaying $W$ and 
not in the lab-frame. 

The $D$ functions can be decomposed into the sum of a symmetric and  
an antisymmetric part under the simultaneous exchange 
$\theta \leftrightarrow \pi-\theta$ and 
$\phi \leftrightarrow \phi + \pi$ \cite{gounaris}. 
When we discuss the 
phenomenology of hadronic
$W$ decays in section 3 
we will consider only the symmetric pieces of the 
$D$ functions, because in 
hadronic W--decays, quark, antiquark and gluon jets cannot 
be distinguished and therefore the antisymmetric parts 
of the $D$ functions drop out. 
In this section, however, we shall present complete results 
both for the symmetric and the antisymmetric terms of the 
W decay functions. 

Our aim is to calculate the one loop QCD corrections to the 
decay functions $D_{AB}$. If gluon emission is taken into account, 
these functions depend on three more variables $\chi, x_1$ and 
$x_2$ in addition to $\theta$ and $\phi$, all defined in the 
$W^-$ rest frame, as follows:  
$x_1$ and $x_2$ are the rescaled energies of the quark q (with 
4--momentum $p_1$) and the antiquark q' (with 4--momentum $p_2$). 
They can be given in terms of invariant dot products as  
\begin{equation}
2p_2p_3=m_W^2(1-x_1) \qquad 
2p_1p_3=m_W^2(1-x_2) \qquad
2p_1p_2=m_W^2(1-x_3) 
\label{eq6}                      
\end{equation}
$x_3$ is the rescaled energy of the gluon, with momentum 
$p_3$ and with $x_3=2-x_2-x_1$ from 
energy conservation.
$\theta$, $\phi$ and $\chi$ are the angles needed to fix the 
spatial orientation of the 
triangle which is formed by the vectors 
$\vec p_1$, $\vec p_2$ and $\vec p_3$ 
in the $W^-$ rest frame (in lowest order 
there are only two angles $\theta$ and $\phi$ necessary to fix the 
quark(=--antiquark) direction in the $W^-$ rest frame).
After integration over the additional variables $\chi$, 
$x_1$ and $x_2$, the form 
of the cross section eq. \ref{eq1} will be left unchanged, with 
\begin{equation}
D_{AB}(\theta ,\phi)=D_{AB}^0(\theta ,\phi)
+C_F{\alpha_s \over 2\pi}D_{AB}^1(\theta ,\phi) 
\label{eq7}                      
\end{equation} 
replacing $D_{AB}^0$ in eq. \ref{eq1}. 
$D_{AB}^1$ is the sum of all one loop QCD corrections 
(virtual, soft and hard gluons). 
We shall actually integrate in such a way that the triangle is 
'reduced' to the thrust direction, where the thrust is defined in the 
ordinary way except that it is determined in the W rest frame. 
For example, for three partonic jets, q ,$\bar q'$ and g, the thrust 
direction is just the direction of the most energetic parton. 
This definition is useful, because experimentally quark, 
antiquark and gluon jets cannot be distinguished. The transformation 
to the $W$ rest frame should be no problem, too, because the 
hadronic part of the event in general can be completely reconstructed. 
 
The calculation of the $D_{AB}^1(\theta ,\phi)$ will be described 
in the following. More precisely, we shall calculate the correction 
to the ratio ${D_{AB} \over D_{total}}$, because in this ratio all 
ultraviolet, infrared and collinear singularities present in the 
higher order matrix element drop out. 
This is a consequence of the universality of those 
singularities, which has been known to be true in QCD for many 
years. Otherwise such singularities 
would appear in intermediate steps of the calculation. For example, 
one has 
\begin{equation} 
D_{total}(q \bar q' g)=C_F{\alpha_s \over 2\pi} 
{x_1^2+x_2^2 \over (1-x_1)(1-x_2)}D_{total}^0
\label{eq8}
\end{equation}
with singularities for $x_{1,2} \rightarrow 1$. Those ratios 
have the further advantage that the virtual gluon exchange corrections 
drop out (for the case of massless quarks which we assume throughout). 
Writing 
\begin{equation}
D_{total}=D_{total}^0+C_F{\alpha_s \over 2\pi} D_{total}^1
=2(1+{\alpha_s \over \pi})
\label{eq9}
\end{equation}
(defined including virtual gluon exchange) we have 
\begin{equation}
{D_{AB} \over D_{total}}={D_{AB}^0 \over 2}
(1+C_F{\alpha_s \over 2\pi}{X_{AB}\over 2D_{AB}^0}) 
\label{eq10}
\end{equation}
with 
\begin{equation}
\begin{split}
X_{AB}:&=2D_{AB}^1-D_{AB}^0D_{total}^1 \\ 
&=\int_0^{2\pi}{d\chi \over 2\pi}\int_0^1 dx_1 \int_{1-x_1}^1 dx_2 
H_{\mu\nu}(q \bar q' g) (2 \epsilon_A^{\mu}\epsilon_B^{\ast\nu} 
-D_{AB}^0(-g_{\mu\nu}+{W_{\mu}W_{\nu}\over m_W^2})) 
\label{eq11}
\end{split}
\end{equation}
What is the form of the $W^-$ polarization vectors $\epsilon_A^{\mu}$? 
In the 
actual calculation we worked in the $W^-$ rest system and we choose  
the quantization axis to be given by 
the direction of the quark momentum, $q_1={m_W \over 2}x_1(1,0,0,1)$ 
(or, more generally, the thrust direction).  
Furthermore, we defined the coordinate system such that 
$q_2={m_W \over 2}x_1(1,0,\sin \theta_{12},\cos\theta_{12})$ 
is the antiquark momentum, 
where $\cos\theta_{12}$ is given by 
$1+{2\over x_1x_2}-{2\over x_1}-{2\over x_2}$ 
simply from energy--momentum conservation. 
Therefore we had to produce the polarization vectors from the 
ordinary vectors $(0,0,0,-1)$ and ${1\over\sqrt{2}}(0,\pm 1,i,0)$ 
by an arbitrary Euler rotation given in terms of $\phi$, $\theta$ 
and $\chi$, 
\begin{equation} 
\epsilon_L=(0,-\sin\theta\cos\chi,-\sin\theta\sin\chi,-\cos\theta)
\label{eq12}
\end{equation}
\begin{equation} 
\epsilon_{\pm}={e^{\pm i\phi} \over\sqrt{2}} 
(0,-i\cos\chi\mp\cos\theta\sin\chi,
-i\sin\chi\pm\cos\theta\cos\chi,\pm\sin\theta)
\label{eq13}
\end{equation}
Note that the angles $\chi$ and $\theta$ 
enter the angular correlation structure of the cross section 
for $Z \rightarrow q \bar q g$ as well. 
The generic form of the lowest order LEP1 cross section is an 
expansion in powers of $\cos \theta$  
\begin{equation} 
{d\sigma(e^+e^- \rightarrow Z \rightarrow q \bar q) \over d\cos \theta} 
={3 \over 8} \sigma_U^0 (1+\cos^2\theta)
+{3 \over 4}\sigma_L^0 \sin^2\theta +
{3 \over 4}\sigma_P^0 \cos \theta 
\label{eqlo}
\end{equation}  
where the last term is the parity violating vector--axialvector interference 
term, which cannot be measured, because quarks and antiquarks cannot 
be distinguished.   
For a three jet process one has an expansion  
\begin{multline} 
{2 \pi d\sigma(e^+e^- \rightarrow Z \rightarrow q \bar q g) 
\over d\cos \theta d\chi dx_1 dx_2} 
={3 \over 8} {d\sigma_U(x_1,x_2) \over dx_1dx_2} (1+\cos^2\theta)
+{3 \over 4}{d\sigma_L(x_1,x_2) \over dx_1dx_2} \sin^2\theta  \\ 
+{3 \over 4}{d\sigma_P(x_1,x_2) \over dx_1dx_2} \cos \theta 
-{3 \over 2 \sqrt{2}}{d\sigma_N(x_1,x_2) \over dx_1dx_2}\sin 2\theta \cos \chi
\\
+{3 \over 4}{d\sigma_T(x_1,x_2) \over dx_1dx_2} \cos 2\chi \sin^2 \theta 
-{3 \over 2 \sqrt{2}}{d\sigma_M(x_1,x_2) \over dx_1dx_2}\sin \theta \cos \chi
\end{multline}  
with ${d\sigma_L(x_1,x_2) \over dx_1dx_2}=
2{d\sigma_T(x_1,x_2) \over dx_1dx_2}$ ( as long as one 
neglects higher orders, i.e. two gluon emission). 
If one integrates over $\chi$, one recovers the structure of the 
lowest order eq. \ref{eqlo}. 
In that case, for normalized distributions, there 
are only two independent quantities which fix the cross section 
completely, namely 
${\sigma_L\over\sigma_{total}}=-{\sigma_U\over\sigma_{total}}$ 
and ${\sigma_P\over\sigma_{total}}$. These quantities, after 
integration, correspond to L and P eqs. \ref{eql} and \ref{eqp}, 
and have been introduced in a different context in the 
last two references of \cite{lampe}. 
In hadronic Z decays 
the parity violating piece  
${\sigma_P\over\sigma_{total}}$ cannot be determined, because 
quark and antiquark jets cannot be distinguished. 

In our case we shall find matrix elements which depend  
on all three angles $\phi$, $\chi$ and $\theta$ 
and have to integrate over $\chi$ to get the $D_{AB}^1$, eq. 
\ref{eq7}. The appearance of an additional angle is due to 
the presence of an additional plane, the plane spanned by 
the momenta of $e^{\pm}$ and $W^{\pm}$.   

In the difference 
$X_{AB}$ all singularities and contributions 
from virtual gluon exchange drop out. For example, for 
A=B=L one has 
\begin{multline} 
{D_{LL} \over D_{total}}={\sin^2\theta \over 2}+ 
C_F{\alpha_s \over 2\pi}
\int_0^{2\pi}{d\chi \over 2\pi}\int dx_1 \int dx_2  
\Bigl\{ 2{x_1+x_2-1 \over x_1^2}(1-2\sin^2\theta \\ 
+\cos^2\chi\sin^2\theta)
-(2{x_2 \over x_1}-{x_2^2 \over (1-x_1)(1-x_2)})\sin\theta_{12} 
\sin\chi\sin\theta\cos\theta \Bigr\} 
\label{eq14}
\end{multline} 
in case that the thrust (=quantization) axis is given by $\vec p_1$, 
i.e. $x_1 > x_{2,3}$. Note that ${\sin\theta_{12}\over (1-x_1)(1-x_2)}$ 
is integrable. 
Analogous results are obtained in the other two cases, 
$x_2 > x_{1,3}$ and $x_3 > x_{1,2}$. 
Integrating these results over the appropriate 
phase space regions \cite{kramer} one obtains 
\begin{equation} 
{D_{LL} \over D_{total}}={\sin^2\theta \over 2}
(1-3LC_F{\alpha_s \over2\pi})+LC_F{\alpha_s \over 2\pi}
\label{eq15}
\end{equation}
where 
\begin{equation} 
L=0.4875 
\label{eql}
\end{equation}
is the numerical result of the integration over 
$x_1$ and $x_2$. Similarly, for the other decay functions:  
\begin{equation} 
{D_{++}+D_{--} \over D_{total}}={1+\cos^2\theta \over 2}
(1-3LC_F{\alpha_s \over2\pi})+2LC_F{\alpha_s \over 2\pi}
\label{eq16}  
\end{equation}
\begin{equation} 
{D_{++}-D_{--} \over D_{total}}=-\cos\theta
(1+PC_F{\alpha_s \over2\pi})
\label{eq17}  
\end{equation}
where 
\begin{equation} 
P=-1.340 
\label{eqp}
\end{equation}
is the characteristic correction for a parity violating 
contribution as can be derived from eq. \ref{a3} of the appendix. 
This and various details concerning the matrix elements will be 
discussed in the appendix. 
For the phenomenological applications we have in mind here, 
it is not relevant, because 
quark and antiquark jets cannot be distinguished and 
therefore the parity violating
contribution cannot be determined experimentally. Note that, in addition, 
there are other 
contributions from the matrix elements, which  
disappear when the integration over $\chi$ is performed 
(see the appendix). 

Next, we find   
\begin{equation} 
{D_{+-} \over D_{total}}=e^{2i\phi}{\sin^2\theta \over 4}
(1-3LC_F{\alpha_s \over2\pi})
\label{eq18}  
\end{equation}
\begin{equation} 
{D_{\pm L} \over D_{total}}={e^{\pm i\phi}\over2\sqrt{2}}
\Bigl\{\pm\sin\theta\cos\theta(1-3LC_F{\alpha_s \over2\pi}) 
-\sin\theta (1+PC_F{\alpha_s \over2\pi}) \Bigr\}
\label{eq19}  
\end{equation}
Note that we know $D_{total}$ from eq. \ref{eq9}. In any 
case, the ratios ${D_{AB} \over D_{total}}$ are most interesting,  
because they enter in normalized distributions, which are more 
sensitive to anomalous triple gauge boson couplings than the total 
cross section. 

Forgetting about the parity violating contributions, 
our results eqs. \ref{eq15} -- \ref{eq19} can be brought to a 
very compact form by defining  
\begin{equation}
\hat{D}_{AB}={1 \over 1+{\alpha_s \over \pi}}{D_{AB}
\over 1-3LC_F{\alpha_s \over 2\pi}}
\label{eq21}
\end{equation}
Using this, eqs. \ref{eq15} -- \ref{eq19}, 
can be rewritten as
\begin{equation}
\hat{D}_{AB}=D^0_{AB}+2\delta_{AB}LC_F{\alpha_s \over 2\pi}
+O(\alpha_s^2)
\label{eq20}
\end{equation}
This is the central result of this article. 
It means that the QCD corrections can be organized such that 
apart from an overall normalization factor 
only the diagonal decay functions $D_{LL}$, $D_{++}$ and 
$D_{--}$ are modified. They 
are modified by a constant term $2LC_F{\alpha_s \over 2\pi}$.  
The absolute numerical value of 
$2LC_F{\alpha_s(m_W^2) \over 2\pi}=0.024$ is relatively 
small, but may be relevant in regions, in which the lowest order terms 
vanish, like $D_{LL}^0$ at $\theta=0$. 
To see the significance of the term $2LC_F{\alpha_s \over 2\pi}$ 
one may also consider 
$\hat{D}_{++}$. The constant correction may be written as
$\sim \cos^2\theta + \sin^2\theta$. Comparing with the 
lowest order decay function $D_{++}^0$, eq. \ref{eq3},  
we see that a purely transverse state appears to accquire 
a longitudinal component, thus changing the relative proportion of 
longitudinal and transverse polarisations in the final state. Such 
effects are among the hallmarks 
of non-standard triple gauge boson couplings.   
Although the absolute numerical value of 
$2LC_F{\alpha_s(m_W^2) \over 2\pi}$ is relatively 
small, it is certainly comparable in size with 
possible non-standard effects, which may be even smaller.  
Concrete phenomenological applications will be presented in the 
next section. 

\vskip2cm

{\bf 3. Phenomenological Applications} 

Although the QCD corrections are remarkably simple, they modify the 
shapes of all the distributions relevant for the analysis of the 
triple gauge boson vertex, 
like azimuthal or polar angle distributions of one or both W's. We will
present both analytical and numerical results
for these distributions. Our basic input parameters will 
be $\sqrt{s} = 190$ GeV for the total beam energy and 
$\alpha_s =0.11$ for the magnitude of the strong coupling 
constant. The latter value is reasonable in view of the 
LEP1 result for $\alpha_s$ 
and the uncertainty in the renormalization scale. 
(The argument of $\alpha_s$ should be somewhere between 
$m_W$ and $\sqrt{s}$.)  
We shall present all results in the form of a ratio of 
higher order to lowest order prediction, because this emphasizes  
the QCD effects and reduces side effects like the uncertainty 
in $m_{W}$ etc. 

Let us first discuss the case, in which only one of the two
W decays is considered, 
i.e. $e^+e^- \rightarrow W^+W^- \rightarrow  W^+ q \bar q'$. 
The azimuthal differential distribution 
may be written as  \cite{gounaris}
\begin{equation}
{d \sigma(e^+e^- \rightarrow W^+j_-X)
\over d\cos\vartheta d\cos\theta_- d\phi_-}=
{3 \over 8\pi}{d \sigma(e^+e^- \rightarrow W^+W^-)
\over d\cos\vartheta} \sum_{A,B} \rho_{AB} D_{AB}
\label{eq22}
\end{equation}
where $\rho_{A,B}$ is the spin density matrix of the $W^-$ as given, 
for example, in \cite{gounaris}. $j_-$ denotes the highest 
energetic jet from 
the $W_-$ decay used to define the thrust direction. 
The angles $\theta$ and $\phi$ defining the thrust direction 
have been renamed 
to $\theta_-$ and $\phi_-$ to stress that they refer to the 
$W^-$ decay. Note that for hadronic W--decays it is usually 
possible to transform to the W rest system, because 
the W--momentum can be fully reconstructed from the 
momenta of the decay products.  
Using our result eq. \ref{eq20} one can rewrite eq. 
 \ref{eq22} as
\begin{multline}
{d \sigma(e^+e^- \rightarrow W^+q\bar{q})
\over d\cos\vartheta d\cos\theta_- d\phi_-}=
{3 \over 8\pi}{d \sigma(e^+e^- \rightarrow W^+W^-)
\over d\cos\vartheta}(1+{\alpha_s \over \pi})
(1-3LC_F{\alpha_s \over 2\pi}) \\ 
\times 
\Bigl\{\sum_{A,B} \rho_{AB} (D^0_{AB}+
2LC_F{\alpha_s \over 2\pi}\delta_{AB}) \Bigr\}
\label{eq23}
\end{multline}
Carrying out the polar angle integrations one obtains 
\begin{equation}
{d \sigma(e^+e^- \rightarrow W^+j_-X)
\over d\phi_-} \sim {4\over 3}+4LC_F{\alpha_s \over 2\pi} 
+{4\over 3}(Re \; r_{+-} \cos 2\phi_- 
-  Im \; r_{+-}\sin 2\phi_- ) 
\label{eq24}
\end{equation}
for the shape of the azimuthal distribution.  
Here $r_{+-}$ is the weighted average of $\rho_{+-}$, 
\begin{equation}  
r_{+-}={\int d \cos \vartheta {d\sigma \over d \cos \vartheta} \rho_{+-} 
\over \int d \cos \vartheta {d\sigma \over d \cos \vartheta}}
\label{eq25}        
\end{equation}    
In fig. 1 the ratio ${{d \sigma \over d\phi_-}\vert _{ho} 
\over {d \sigma \over d\phi_-}\vert _{lo}}$ is shown 
using the standard model values for $r_{+-}$  
(Im $r_{+-}=0$ and Re $r_{+-}=-0.53$). 
Nonstandard triple gauge boson vertex interactions modify 
the $\phi_-$ (as well as other) 
distributions, by modifying $r$ and $\rho$, respectively. 
For example, if there are CP violating gauge boson couplings, 
$r$ and $\rho$ may acquire non-vanishing imaginary pieces, apart
from modifications of the standard model predictions for the real pieces
induced by all nonstandard effects whether CP violating or not.

In fig. 1, $\vartheta$ and $\theta_-$ have been averaged out completely. 
Interesting information may also be gained by retaining some of the 
dependence in the polar angles. 
For example, 
one may consider the $\phi$ dependence of the forward-backward 
asymmetry in $\vartheta$, in order to keep 
track of the parity violating terms ($\sim\cos\vartheta$) in the 
triple gauge boson interaction. 
The ratio of this distribution for ho and lo is also 
included in fig. 1.
We conclude that the QCD effects must be taken  
into account, if one wants to check the standard model triple gauge boson 
vertex to an accuracy of 1$\%$. 
 
\begin{figure}
\begin{center}
\epsfig{file=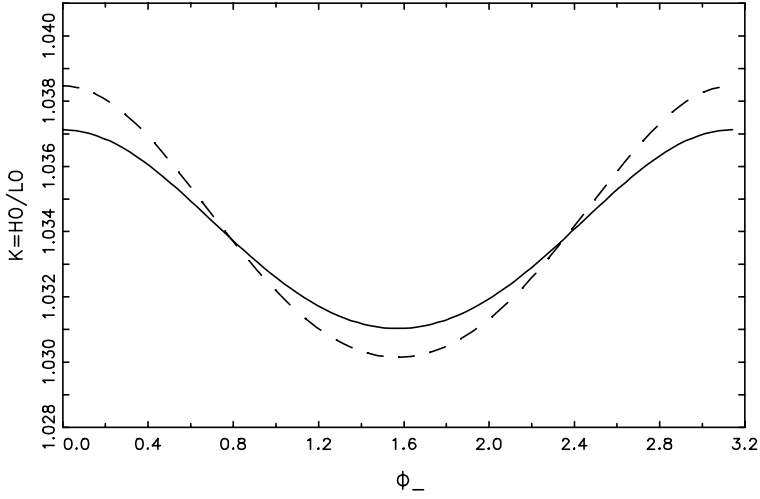,height=10cm,angle=270}
\bigskip
\caption{The ratio ${{d \sigma \over d\phi_-}\vert _{ho}
\over {d \sigma \over d\phi_-}\vert _{lo}}$ as a function of 
$\phi_-$. The dependence on $\theta_-$ and 
$\vartheta$ have been averaged out (full curve). 
To obtain the dashed curved we have integrated over 
$\cos \vartheta$ in an antisymmetric way, i.e. the integrand 
is counted negative, if $\cos \vartheta$ is negative.}
\end{center}
\end{figure}

Furthermore, one may examine distributions, in which $\phi_-$ instead 
of $\theta_-$ is integrated out, 
\begin{multline}
{d \sigma(e^+e^- \rightarrow W^+j_-X)
\over d\cos \vartheta d\cos\theta_-} ={3\over 4} 
{d \sigma(e^+e^- \rightarrow W^+W^-)\over d\cos\vartheta}
\Bigl\{
(\rho_{++}+\rho_{--}) 
\\ \times
\bigl\{{1\over 2}(1+\cos^2\theta_-)+2LC_F{\alpha_s \over 2\pi}\bigr\}
+\rho_{LL}\bigl\{\sin^2\theta_-+2LC_F{\alpha_s \over 2\pi}\bigr\}
\Bigr\}
(1+{\alpha_s \over \pi})
(1-3LC_F{\alpha_s \over 2\pi})
\label{eq26}
\end{multline}
In this equation the parity violating piece $\sim \cos\theta_-$ 
has been left out, because it cannot be seen in the hadronic 
$W^-$ decay. Since the relative size of QCD corrections is not the
same for the transverse and longitudinal part of eq. \ref{eq26} 
the shape of the $\theta_-$ distribution is modified 
by the QCD term. In fig. 2 the ratio of the ho to lo distribution 
is shown as a function of $\theta_-$, both for symmetric 
and antisymmetric integration over $\vartheta$.  
Again, the quantitative results contained in this figure 
will allow to discriminate QCD from nonstandard effects. 
Clearly, the variation of the curves in fig. 2 must be taken 
into account, if one wants to check the standard model triple gauge boson
vertex to an accuracy of 1$\%$. 
Note that in all the figs. 1, 2 and 3 below, the average 
of the curves correspond to the overall QCD correction 
factor $1+{\alpha_s \over \pi}$. 

\begin{figure}
\begin{center}
\epsfig{file=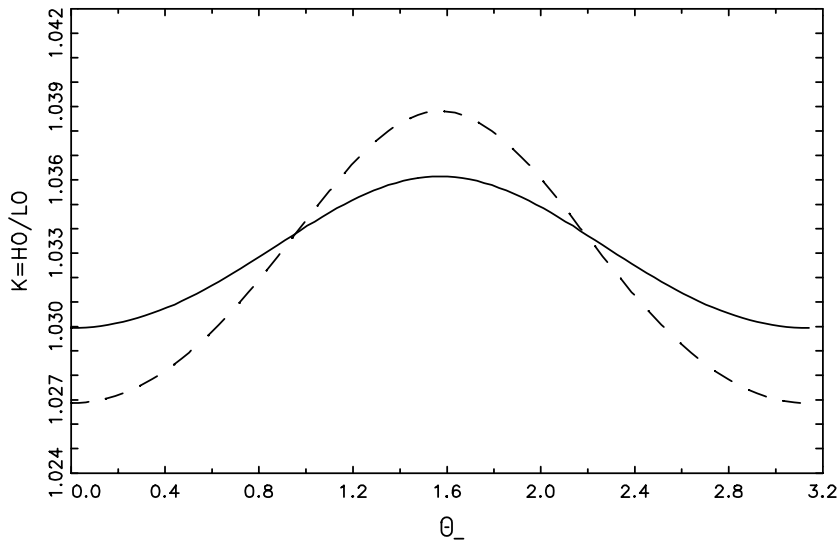,height=11cm,angle=270}
\bigskip
\caption{
The ratio of the ho to lo  
as a function of $\theta_-$, both for symmetric 
and antisymmetric integration over $\vartheta$.}
\end{center}
\end{figure}

If the decay of the $W^+$ is taken into account and 
both W's decay hadronically, the situation becomes more complicated.  
For example, the shape of the double differential azimuthal distribution 
is of the generic form  
\begin{multline}            
{d \sigma(e^+e^- \rightarrow W^+W^- \rightarrow j_+j_-X)
\over d\phi_+ d\phi_-} \sim 
\sum_{ABA'B'} r_{ABA'B'}D_{AB}^0D_{A'B'}^0
\\
+2LC_F{\alpha_s \over 2\pi} 
\sum_{A'B'}r_{A'B'}D_{A'B'}^0 
+2LC_F{\alpha_s \over 2\pi} 
\sum_{AB}r_{AB}D_{AB}^0 
\label{eq27}                               
\end{multline} 
where $r_{ABA'B'}$, $r_{AB}$ and $r_{A'B'}$ are defined in 
analogy with $r_{+-}$, eq. \ref{eq25}. 
The angles $\phi_-$ and $\phi_+$ refer to the decay of 
the $W^-$ and $W^+$, respectively. 
If only one of the W's decays hadronically one obtains a 
similar result for ${d \sigma \over d \phi_{+} d \phi_{-}}$, 
with only one term $\sim 2LC_F$... instead 
of two such terms. Note that in eq. \ref{eq1} $\phi_-$ and $\phi_+$ 
were denoted by $\phi$ and $\phi_l$, respectively. 
We present the numerical results on the effect of the higher order 
corrections on the double differential azimuthal 
distribution $\frac{d \sigma}{d \phi_{-} d \phi_{+}}$ 
in table 2 (for the case that one W decays leptonically) and 
table 3 (for the case that both W's decay hadronically). 
The entries in the table are 
the ratio of higher order
to leading order cross-sections in the respective $\phi_{+}$--$\phi_{-}$ 
bins. 
All angles are in units of $\pi$.
We see that the variation of the numbers in table 2 is en gross about 
the same as the variations of the curves in figs. 1 and 2, and 
conclude that in all the cases it is necessary to take these 
QCD corrections into account, if one wants to pin down 
the standard model triple gauge boson vertex to an accuracy of 1-2$\%$. 
Of course, the overall QCD K factor $1+{\alpha_s \over \pi}$ 
was known before, but in figs. 1, 2 and 3 and tables 2 and 3 we see 
that it is also important to know the deviations from this 
average.  

Conversely, if one leaves the polar angle dependence and 
integrates over the azimuthal angles, one obtains 
\begin{multline}
{d \sigma(e^+e^- \rightarrow j_+j_-X)
\over d\cos \vartheta d\cos\theta_-d\cos\theta_+} =({3\over 4})^2
{d \sigma(e^+e^- \rightarrow W^+W^-)\over d\cos\vartheta}
\Bigl\{
(\rho_{++++}+\rho_{++--}\rho_{--++}+\rho_{----})
\\ \times
\bigl\{{1\over 2}(1+\cos^2\theta_-)+2LC_F{\alpha_s \over 2\pi}\bigr\}
\bigl\{{1\over 2}(1+\cos^2\theta_+)+2LC_F{\alpha_s \over 2\pi}\bigr\}
\\
+\rho_{LLLL}
\bigl\{\sin^2\theta_-+2LC_F{\alpha_s \over 2\pi}\bigr\}
\bigl\{\sin^2\theta_++2LC_F{\alpha_s \over 2\pi}\bigr\}
\\
+(\rho_{LL++}+\rho_{LL--})
\bigl\{\sin^2\theta_-+2LC_F{\alpha_s \over 2\pi}\bigr\}
\bigl\{{1\over 2}(1+\cos^2\theta_+)+2LC_F{\alpha_s \over 2\pi}\bigr\}
\\
+(\rho_{++LL}+\rho_{--LL})
\bigl\{\sin^2\theta_++2LC_F{\alpha_s \over 2\pi}\bigr\}
\bigl\{{1\over 2}(1+\cos^2\theta_-)+2LC_F{\alpha_s \over 2\pi}\bigr\}
\Bigr\}
\\ \times
(1+{\alpha_s \over \pi})^2
(1-3LC_F{\alpha_s \over 2\pi})^2
\label{eq28}
\end{multline}
where $\rho_{ABA'B'}(s,\cos \vartheta)$ denotes 
the two-particle joint density matrix for $W^{\pm}$ production 
as defined in ref. \cite{gounaris}. 
The complicated looking formulae eqs. \ref{eq27} and 
\ref{eq28} can be verified easily by 
combining the lowest order formulae of reference \cite{gounaris}
with the compact form of our higher order result eq. \ref{eq20}.  
Terms $\sim \cos\theta_+$ and $\sim \cos\theta_-$ 
have been left out, because it has been assumed that both W's
decay hadronically and those terms cannot be detected in that case. 

This case of both W's decaying hadronically poses additional 
QCD problems, which we will not discuss, but 
which have recently become the 
focus of some interest \cite{khoze} 
\cite{kleiss}. These have to do with finite width 
effects, which are partially nonperturbative and thus cannot be 
calculated from first principles, with important consequences for the 
precision to which the Standard Model can be tested at LEP200.

The distributions eq.  
\ref{eq28} may also be considered in case that 
only one W (e.g. the $W^-$) decays 
hadronically. In that case in eq. \ref{eq28}  
one has to add the parity violating pieces $\sim\cos \theta_+$, 
because they can be determined from the direction of the $l^+$. 
\begin{multline}
{d \sigma(e^+e^- \rightarrow l^+\nu j_-X)
\over d\cos \vartheta d\cos\theta_-d\cos\theta_+} =({3\over 4})^2
{d \sigma(e^+e^- \rightarrow W^+W^-)\over d\cos\vartheta}
\Bigl\{
(\rho_{++++}+\rho_{++--}+\rho_{--++}+\rho_{----})
\\ \times
\bigl\{{1\over 2}(1+\cos^2\theta_-)+2LC_F{\alpha_s \over 2\pi}\bigr\}
{1\over 2}(1+\cos^2\theta_+)
\\
+\rho_{LLLL}  
\bigl\{\sin^2\theta_-+2LC_F{\alpha_s \over 2\pi}\bigr\}
\sin^2\theta_+
\\
+(\rho_{LL++}+\rho_{LL--})
\bigl\{\sin^2\theta_-+2LC_F{\alpha_s \over 2\pi}\bigr\}
{1\over 2}(1+\cos^2\theta_+)
\\
+(\rho_{++LL}+\rho_{--LL})
\sin^2\theta_+
\bigl\{{1\over 2}(1+\cos^2\theta_-)+2LC_F{\alpha_s \over 2\pi}\bigr\}
\\
-(\rho_{++++}-\rho_{++--}+\rho_{--++}-\rho_{----})
\bigl\{{1\over 2}(1+\cos^2\theta_-)+2LC_F{\alpha_s \over 2\pi}\bigr\}
\cos\theta_+
\\
-(\rho_{LL++}-\rho_{LL--})
\bigl\{\sin^2\theta_-+2LC_F{\alpha_s \over 2\pi}\bigr\}
\cos\theta_+
\Bigr\}
\\ \times
(1+{\alpha_s \over \pi})
(1-3LC_F{\alpha_s \over 2\pi})
\label{eq29}
\end{multline}
Note that now the factors depending on $\theta_+$  
do not get the QCD term. 
In fig. 3 the ratio of the ho to the lo of the antisymmetric 
piece in $\cos \theta_+$ (coefficient of $\cos \theta_+$ in 
the last equation) is shown as a function of $\theta_-$, 
both for symmetric and antisymmetric integration in $\vartheta$. 
The variation of the curves is somewhat larger than in figs. 1 and 2, 
so that essentially the same conclusions can be drawn.  

\begin{figure}
\begin{center}
\epsfig{file=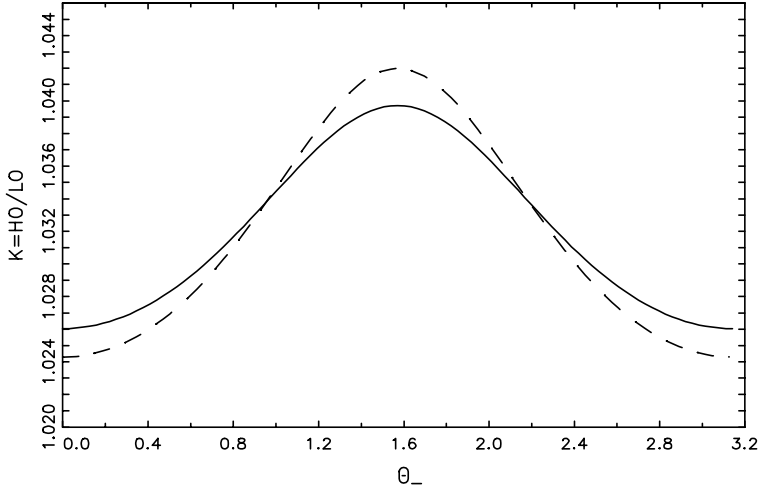,height=10cm,angle=270}
\bigskip
\caption{
The ratio of the ho to the lo of the antisymmetric 
piece in $\cos \theta_+$ (coefficient of $\cos \theta_+$ in
eq. \ref{eq29}) as a function of $\theta_-$,
both for symmetric (solid curve) and antisymmetric (dashed curve) 
integration in $\vartheta$.}
\end{center}
\end{figure}

\vskip2cm

{\bf Conclusions}

In this article we have presented a complete calculation of 
one loop QCD corrections to W--pair production and decay at LEP200, 
including W--polarization effects. 
We have put special emphasis on the effects on those angular 
distributions sensitive to the structure of  
triple gauge boson vertices. Our results are of the order of a few percent 
and thus are certainly as important as finite W--width effects and must be 
taken into account for a precision study of the pure gauge sector of the 
standard model.

This work is part of a larger project, which one of the authors 
has pursued over the last years, namely to calculate QCD corrections 
to differential distributions using the known QCD effects on total 
rates plus the real gluon matrix elements \cite{lampe}. 
This method has been applied successfully even to two loop 
problems \cite{lampe} at LEP1. It could certainly be used  
to extend eq. \ref{eq20} to two loops. We have not attempted this, because 
it is unlikey to be relevant given 
the magnitude of statistical and systematic errors expected at LEP200.

\vskip2cm

{\bf Acknowledgements} 

We are indebted to several colleagues who helped us in the course 
of this work. First of all, the idea to study the problem arose 
in 
discussions with D. Zeppenfeld. Secondly, G. Kramer helped us to 
understand the relation between QCD corrections to Z production at LEP1  
and W--pair production at LEP200. Finally, 
we thank J.-L. Kneur for a program that allowed us to check
our lowest order expressions. 
K.J.A wishes to acknowledge the
hospitality of the Max Planck Institut where this work was commenced.

\vskip2cm


{\bf Appendix: The complete W decay functions to O($\alpha_s$)}

Before integration over $\chi$, $x_1$ and $x_2$ 
one has a cross section 
\begin{displaymath}
{d\sigma(e^+e^- \rightarrow W^+W^- \rightarrow l^+ \nu_l
q \bar q' g) \over d\cos \vartheta d\cos \theta_l d\phi_l
d\cos \theta d\phi dx_1 dx_2 d\chi} \end{displaymath}
containing decay functions 
$D_{AB}(\theta,\phi,\chi,x_1,x_2)$ which depend on $\chi$, 
$x_1$ and $x_2$ in addition to $\theta$ and $\phi$. 
These will be given in the following. First, for A=B=L, one has
\begin{multline}
{D_{LL}(\theta,\phi,\chi,x_1,x_2) \over D_{total}}= 
{\sin^2\theta \over 2}\delta(1-x_1)\delta(1-x_2)\delta(\chi)+    
C_F{\alpha_s \over 2\pi}
\Bigl\{ l(x_1,x_2)(1-2\sin^2\theta \\
+\cos^2\chi\sin^2\theta)
+m(x_1,x_2)    
\sin\chi\sin\theta\cos\theta \Bigr\}
\label{a1}
\end{multline}
where $D_{total}=2(1+{\alpha_s \over \pi})$. 
The functions $l(x_1,x_2)$ and $m(x_1,x_2)$ 
are shown in table 1. The integral of $l$ over 
$dx_1dx_2$ is the number L=0.4875 
as given in the main text eq. \ref{eql}. The integral M of $m$ over
$dx_1dx_2$ is given in table 1.  
$m$ does not contribute to 
${d\sigma \over d\cos \vartheta d\cos \theta_l d\phi_l
d\cos \theta d\phi}$ 
because the coefficient of $m$ vanishes 
when the integral over $\chi$ is performed. It contributes only 
to the azimuthal (=$\chi$) dependence in 3-jet decays. 
Secondly, one finds  
\begin{multline}
{(D_{++}+D_{--})(\theta,\phi,\chi,x_1,x_2) \over D_{total}}=
{1+\cos^2\theta \over 2}\delta(1-x_1)\delta(1-x_2)\delta(\chi)
\\ -    
C_F{\alpha_s \over 2\pi}
\Bigl\{ m(x_1,x_2) \sin\chi\sin\theta\cos\theta   
+l(x_1,x_2)                                                    
(1-2\sin^2\theta+\cos^2\chi\sin^2\theta) \Bigr\}
\label{a2}
\end{multline}            

\begin{multline}
{(D_{++}-D_{--})(\theta,\phi,\chi,x_1,x_2) \over D_{total}}=            
-\cos\theta\delta(1-x_1)\delta(1-x_2)\delta(\chi)
\\ -
C_F{\alpha_s \over 2\pi}
\Bigl\{ p(x_1,x_2) \cos\theta                 
-n(x_1,x_2)
\sin\chi\sin\theta \Bigr\}
\label{a3}
\end{multline}
The function $p(x_1,x_2)$ together with its integral P 
over $x_1$ and $x_2$ is given in table 1. It gives 
a parity violating contribution which is not measurable 
in hadronic W decays. 
The function $n(x_1,x_2)$ and its integral N are given in table 1, too. 
They do
not contribute to the inclusive distributions discussed in the
main text because their coefficient vanishes
when the integral over $\chi$ is performed. 
Furthermore they arise from the parity violating part of the W decay and 
are not measurable
in hadronic W decays.           
Finally, one finds 
\begin{multline}
{D_{+-}(\theta,\phi,\chi,x_1,x_2) \over D_{total}}=            
{e^{2i\phi} \over 2}{\sin^2\theta \over 2}
\delta(1-x_1)\delta(1-x_2)\delta(\chi)
\\ +
{e^{2i\phi} \over 2}C_F{\alpha_s \over 2\pi}
\Bigl\{ l(x_1,x_2) (1-2\cos^2\chi+\sin^2\theta\cos^2\chi \\ 
-2\sin^2\theta+2i\sin\chi\cos\chi\cos\theta)                
-m(x_1,x_2)\sin\theta
(i\cos\chi+\sin\chi\cos\theta) \Bigr\}
\label{a4}
\end{multline}
\begin{multline}
{D_{\pm L}(\theta,\phi,\chi,x_1,x_2) \over D_{total}}=
{e^{\pm i\phi} \over 2\sqrt{2}}( \pm \cos\theta\sin\theta  
-\sin\theta ) 
\delta(1-x_1)\delta(1-x_2)\delta(\chi)
\\ +
{e^{\pm i\phi} \over \sqrt{2}}C_F{\alpha_s \over 2\pi}
\Bigl\{ l(x_1,x_2) \sin\theta(i\cos\chi\sin\chi 
\pm\cos\theta(\cos^2\chi-2))   \\ 
-{1 \over 2} p(x_1,x_2)\sin\theta 
+m(x_1,x_2)
(\pm\sin\chi(\cos^2 \theta -{1\over 2})-{i\over 2}\cos\chi\cos\theta) 
\\
+n(x_1,x_2)(\pm{i\over 2}\cos\chi-{1\over 2}\sin\chi\cos\theta) \Bigr\}
\label{a5}
\end{multline}
Note that $D_{-+}(\theta,\phi,\chi,x_1,x_2)$ and 
$D_{L,\pm}(\theta,\phi,\chi,x_1,x_2)$ can be obtained 
via the relation 
$D_{AB}(\theta ,\phi,\chi,x_1,x_2)=
D^{\ast}_{AB}(\theta ,\phi,,\chi,x_1,x_2)$.

\newpage

\begin{table}
\begin{center}  
\epsfig{file=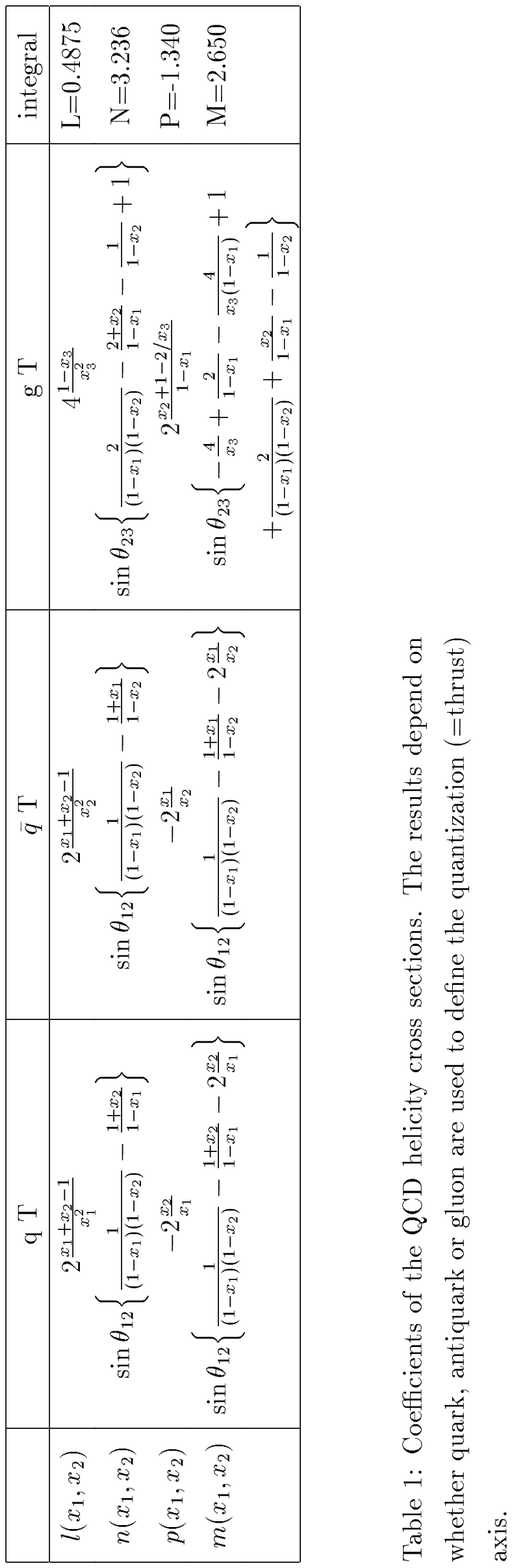,height=25cm,angle=180}
\caption{}
\end{center}
\end{table}

\newpage

\begin{table}
\begin{tabular}{|l || l |l |l |l |l|} \hline
$\downarrow \phi_{+}\|\phi_{-} \rightarrow$ 
& 0 -0.4 & 0.4 -0.8&0.8 -1.2&1.2 -1.6& 1.6 -2.0 \\ \cline{1-6}
0 -0.4 &1.039 &1.029 &1.038 &1.033 &1.031  \\ \cline{1-6}
0.4 -0.8 &1.037 &1.031 &1.034 &1.035 &1.030  \\ \cline{1-6}
0.8 -1.2 &1.034 &1.033 &1.034 &1.033 &1.034  \\ \cline{1-6}
1.2 -1.6 &1.030 &1.035 &1.034 &1.031 &1.037  \\ \cline{1-6}
1.6 -2.0 &1.031 &1.032 &1.038 &1.029 &1.039  \\ \cline{1-6}
\end{tabular}
\bigskip 
\bigskip
\bigskip
\caption{Ratio of ho to lo for the double differential 
azimuthal distribution ${d \sigma \over d \phi_{+} d \phi_{-}}$, 
for various $\phi_{+}$ and 
$\phi_{-}$ bins. $\phi_{+}$ and $\phi_{-}$ are measured in units 
of $\pi$. It is assumed here that the $W^-$ decay is hadronic 
and the $W^+$ decay is leptonic. $\phi_{+}=\phi_{l}$ refers to the 
direction of the charged lepton in the $W^+$ decay. }
\end{table}
\bigskip 
\bigskip
\bigskip
\bigskip 
\bigskip
\bigskip

\begin{table}
\begin{tabular}{|l || l |l |l |l |l|} \hline
$\downarrow \phi_{+}\|\phi_{-} \rightarrow$ 
& 0 -0.2 & 0.2 -0.4&0.4 -0.6&0.6 -0.8& 0.8 -1.0 \\ \cline{1-6}
0 -0.2 &1.076 &1.069 &1.063 &1.064 &1.072  \\ \cline{1-6}
0.2 -0.4 &1.069 &1.068 &1.063 &1.061 &1.064  \\ \cline{1-6}
0.4 -0.6 &1.063 &1.063 &1.063 &1.063 &1.063  \\ \cline{1-6}
0.6 -0.8 &1.064 &1.061 &1.063 &1.068 &1.069  \\ \cline{1-6}
0.8 -1.0 &1.072 &1.064 &1.063 &1.069 &1.076  \\ \cline{1-6}
\end{tabular}
\bigskip 
\bigskip
\bigskip
\caption{Ratio of ho to lo for the double differential 
azimuthal distribution, eq. \ref{eq27}, for various $\phi_{+}$ and 
$\phi_{-}$ bins. 
$\phi_{\pm}$ are the azimuthal angles defined in the decay of 
the $W^{\pm}$ and are measured in units
of $\pi$. 
It is assumed here that both W's decay hadronically. 
Therefore the average here is $(1+{\alpha_s \over \pi})^2 
\approx 1+2{\alpha_s \over \pi}$ 
in contrast to table 2, where the average is 
$1+{\alpha_s \over \pi}$.}
\end{table}

\end{document}